\documentclass[english,manuscript,superscriptaddress]{revtex4-1}
\usepackage[LGR,T1]{fontenc}
\usepackage{color}
\usepackage{babel}
\usepackage{float}
\usepackage{units}
\usepackage{textcomp}
\usepackage{amssymb}
\usepackage{graphicx}
\usepackage{subscript}
\usepackage[unicode=true,pdfusetitle,
 bookmarks=true,bookmarksnumbered=false,bookmarksopen=false,
 breaklinks=false,pdfborder={0 0 1},backref=false,colorlinks=false]
 {hyperref}

\usepackage[resetlabels]{multibib}
\newcites{supp}{Supplemtary References }

\makeatletter

\DeclareRobustCommand{\greektext}{%
  \fontencoding{LGR}\selectfont\def\encodingdefault{LGR}}
\DeclareRobustCommand{\textgreek}[1]{\leavevmode{\greektext #1}}
\ProvideTextCommand{\~}{LGR}[1]{\char126#1}

\makeatother

\begin{document}

\title{Observation of tunable charged exciton polaritons in hybrid monolayer
WS\textsubscript{2} \textendash{} plasmonic nanoantenna system}

\author{Jorge Cuadra }
\email{jorge.cuadra@chalmers.se}

\selectlanguage{english}%

\address{Department of Physics, Chalmers University of Technology, 412 96,
G\"{o}teborg, Sweden.}

\author{Denis G. Baranov\textsuperscript{}}

\address{Department of Physics, Chalmers University of Technology, 412 96,
G\"{o}teborg, Sweden.}

\author{Martin Wers\"{a}ll\textsuperscript{}}

\address{Department of Physics, Chalmers University of Technology, 412 96,
G\"{o}teborg, Sweden.}

\author{Ruggero~Verre}

\address{Department of Physics, Chalmers University of Technology, 412 96,
G\"{o}teborg, Sweden.}

\author{Tomasz J. Antosiewicz}

\address{Department of Physics, Chalmers University of Technology, 412 96,
G\"{o}teborg, Sweden.}

\address{Centre of New Technologies,~University~of~Warsaw,~Banacha 2c,~02-097
Warsaw, Poland.}

\author{Timur Shegai}
\email{timurs@chalmers.se}

\selectlanguage{english}%

\address{Department of Physics, Chalmers University of Technology, 412 96,
G\"{o}teborg, Sweden.}

\begin{abstract}
Formation of dressed light-matter states in optical structures, manifested
as Rabi splitting of the eigen energies of a coupled system, is one
of the key effects in quantum optics. In pursuing this regime with
semiconductors, light is usually made to interact with excitons \textendash{}
electrically neutral quasiparticles of semiconductors, meanwhile interactions
with charged three-particle states \textendash{} trions \textendash{}
have received little attention. Here, we report on strong interaction
between plasmons in silver nanoprisms and charged excitons \textendash{}
trions \textendash{} in monolayer tungsten disulphide (WS\textsubscript{2}).
We show that the plasmon-exciton interactions in this system can be
efficiently tuned by controlling the charged versus neutral exciton
contribution to the coupling process. In particular, we show that
a stable trion state emerges and couples efficiently to the plasmon
resonance at low temperature by forming three bright intermixed plasmon-exciton-trion
polariton states. Our findings open up a possibility to exploit electrically
charged trion polaritons \textendash{} previously unexplored mixed
states of light and matter in nanoscale hybrid plasmonic systems.
\end{abstract}


\maketitle
Interaction between light and matter is at the heart of modern optics.
It is present in atomic physics as well as in solid state systems
and plays an essential role in cavity quantum electrodynamics (cQED).
A regime that attracts significant interest is achieved when the exchange
of energy between photons and matter excitations becomes faster than
the decoherence rates of both subsystems \citep{raimond_manipulating_2001,khitrova_vacuum_2006,smolka_cavity_2014,torma_strong_2015}.
Under these conditions one observes the formation of polaritons that
are hybrid states possessing both light and matter characteristics.
Recently it has been shown that metallic nanoparticles enable this
non-perturbative regime of light-matter interaction due to a combination
of deeply subwavelength confinement of electromagnetic radiation with
large transition dipole moments of molecular and quantum dot excitons
\citep{schlather_near-field_2013,zengin_realizing_2015,santhosh_vacuum_2016,chikkaraddy_single-molecule_2016,wersall_observation_2017}.

Monolayers of transition metal dichalcogenides (TMDCs) are semiconductor
materials with a direct band gap transition \citep{mak_atomically_2010},
where neutral ($X^{0}$) and charged ($X^{+/-}$) excitons (or trions)
are stable and can be excited either by optical or electrical means
\citep{mak_tightly_2013,ross_electrical_2013,ross_electrically_2014}.
The large spin-orbit coupling in these materials permits a spin-valley
degree of freedom accessible by optical dichroism \citep{mak_control_2012,zeng_valley_2012}.
In addition, the broken inversion symmetry puts the band gap at the
edges of the Brillouin zone allowing valley-selective excitations
by pumping with a circularly polarized light. The absorption of TMDCs
monolayers is very high and can reach values of 3\% for MoS\textsubscript{2}
and as much as 10\% for WS\textsubscript{2} at resonance \citep{li_measurement_2014}. 

The high absorptivity of TMDCs monolayers makes them ideal candidates
for realization of the strong coupling regime. Indeed, observations
of strong coupling between TMDCs and nanophotonic resonators have
been reported in several different systems including optical microcavities
\citep{liu_strong_2015,dufferwiel_excitonpolaritons_2015,sidler_fermi_2017}
and diffractive modes of plasmonic nanoparticle arrays \citep{liu_strong_2016,wang_coherent_2016}.
At the single plasmonic nanoparticle level there is no clear evidence
of reaching the strong coupling regime, although efforts have been
made \citep{kern_nanoantenna-enhanced_2015}. Additionally, a number
of recent works report on photoluminescence (PL) modification and
enhancement as a result of plasmon-exciton coupling in the weak coupling
regime \citep{najmaei_plasmonic_2014,butun_enhanced_2015,lee_fano_2015,chen_manipulation_2016,li_tuning_2016}.
It is important to note that, molecular J-aggregates, which have been
extensively used to achieve the strong coupling regime \citep{schlather_near-field_2013,zengin_realizing_2015,wersall_observation_2017},
\citep{bellessa_strong_2004,melnikau_rabi_2016}, suffer from optical
bleaching and are not stable under ambient conditions. In contrast
to that, monolayer TMDCs are chemically inert, stable and robust at
ambient conditions, which makes them advantageous for active and nonlinear
plasmonics applications. 

One of the key features of hybrid plasmon-exciton systems is the possibility
for active control and saturation. Such active control has been demonstrated
in various hybrid nanostructures. Examples include controllable and
reversible switching of the strong coupling regime in microcavities
loaded with photochromic molecules \citep{schwartz_reversible_2011}
and for silver nanoparticle arrays via UV illumination \citep{baudrion_reversible_2013}.
Ultrafast tuning of strongly-coupled metal-molecular aggregates via
femtosecond pumping has also been observed \citep{vasa_ultrafast_2010,vasa_real-time_2013}.
Demonstration of thermalization, cooling and lasing of plasmon-exciton
polaritons have been reported recently \citep{rodriguez_thermalization_2013,vakevainen_plasmonic_2014,ramezani_plasmon-exciton-polariton_2017}.
However, in these works reversible switching and active control was
demonstrated in the systems that involved only electrically neutral
excitations. 

Conversely, little attention has been devoted to strong coupling with
charged excitons. Such interactions would result in the formation
of exciton polaritons carrying a non-zero net electrical charge. These
phenomena have been previously observed in microcavities loaded with
quantum wells \citep{rapaport_negatively_2000,rapaport_charged_2001,rapaport_negatively_2001},
charged quantum dots \citep{rakher_externally_2009} and more recently
in tunable polaron polaritons \citep{sidler_fermi_2017}. The sole
existence of these quasiparticles opens a number of intriguing perspectives,
as such \textquotedblleft charged\textquotedblright{} polaritons are
anticipated to strongly interact with each other and to improve charge
and exciton transport properties mediated by strong coupling \citep{rapaport_negatively_2001,orgiu_conductivity_2015}.

\begin{figure}[H]
\hspace{20mm}\includegraphics[width=120mm]{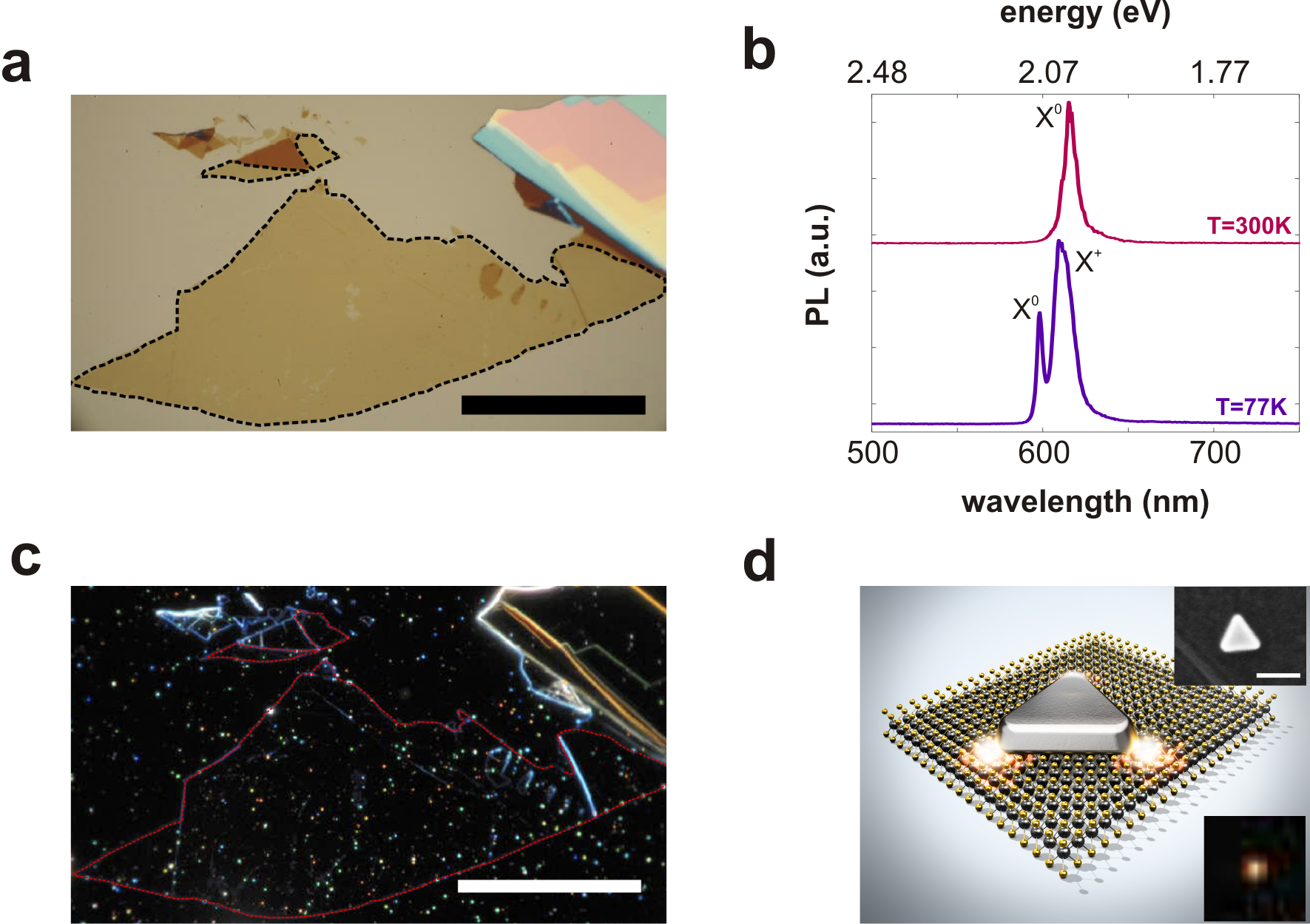}

\caption{\textbf{\footnotesize{}Hybrid 2D material \textendash{} nanoparticle
plasmon system.}{\footnotesize{} (a) Bright field optical image of
the exfoliated WS\protect\textsubscript{\textcolor{black}{\footnotesize{}2}}
flake. Monolayer regions are marked by the dotted line (scale bar
50 \textgreek{m}m) (b) PL spectra of the WS\protect\textsubscript{\textcolor{black}{\footnotesize{}2}}.
The red curve shows the PL taken at ambient conditions, the resonance,
at 616 nm, corresponds to the neutral A-exciton. The blue curve depicts
the PL at 77 K, the peak at 598 nm is the neutral A-exciton that has
blue-shifted. The peak at 611 nm is a charged exciton and dominates
the PL emission. (c) DF microscope image of the WS\protect\textsubscript{\textcolor{black}{\footnotesize{}2}}
flake covered with silver nanoprisms. Monolayer regions are marked
by the dotted line (scale bar 50 \textgreek{m}m). (d) Artist view
of the hybrid system: high density of photonic states (hot-spots)
is shown at the corners of the nanoprism, which overlaps with the
WS}\textcolor{black}{\footnotesize{}\protect\textsubscript{\textcolor{black}{\footnotesize{}2}}}{\footnotesize{}
monolayer for efficient plasmon-exciton interaction. Inset shows the
SEM image of such particle (scale bar 100 nm) and a magnified view
of the DF image.}\textcolor{black}{\footnotesize{}\label{fig:1}}}
\end{figure}

Here, we demonstrate strong interactions between plasmons in an individual
silver nanoprism and neutral as well as charged excitons in monolayer
WS\textsubscript{2}. The latter is especially interesting, as this
opens up a new way to control and manipulate charge via light-matter
interactions. To the best of our knowledge this is the first demonstration
of this kind using an individual plasmonic nanoantenna. In this study
we show that the degree of plasmon-exciton-trion coupling can be tailored
by temperature. In particular, by scanning the temperature in the
range between 77 and 300 K, we are able to observe a transition from
two polaritonic resonances at room temperature, corresponding to plasmon-exciton
interaction, to the formation of three polaritonic resonances at T=77
K, corresponding to both exciton and trion strongly coupled to the
plasmonic cavity.
\\
\\
\textbf{RESULTS}
\\
\textbf{System under study.} The coupled system in this work is composed
of colloidal silver nanoprisms positioned on top of a monolayer WS\textsubscript{2}.
We start by preparing the monolayer structure by mechanical exfoliation
from a high quality crystal and transferring it to a thermally oxidized
silicon substrate. Note that the monolayer flake size can readily
reach sizes greater than \ensuremath{\sim}100 micrometers (see Fig.~\ref{fig:1}a)
The monolayer nature of the WS\textsubscript{2} flake is confirmed
by optical contrast and the bright PL signal \textendash{} typical
for direct bandgap semiconductors (Fig.~\ref{fig:1}b). The PL spectrum
at 300 K under ambient conditions has the resonance at 2.012 eV ($\omega_{X^{0}}$)
and corresponds to the neutral A-exciton with a binging energy of
about 700 meV \citep{zhu_exciton_2015}. The PL signal at 77 K, in
contrast to the room temperature data exhibits two peaks: the high-energy
resonance (2.07 eV) and the low-energy peak (2.03 eV) (blue curve
in Fig.~\ref{fig:1}b). We assign the former to the neutral A-exciton,
which undergoes a blue-shift, whereas the latter corresponds to the
positively charged exciton ($\omega_{X^{+}}$ ) \textendash{} i.e.
a trion. The trion state dominates the PL at low T and has a binding
(or dissociation) energy of about 40 meV in agreement with earlier
reports \citep{zhu_exciton_2015}. The trion state is likely positively
charged \textendash $\omega_{X^{+}}$\textendash{} because of the
positively charged polymer adhesion layer used in this study to bind
plasmonic nanostructures (see Methods and Supplementary Information
(SI)).

The second ingredient to construct the hybrid system in this study
is the silver nanoprism (see Fig. \ref{fig:1}d). These nanoparticles
were prepared by a seed mediated colloidal synthesis and are single
crystalline in nature, which greatly improves the quality factor of
the plasmon resonance and thus the accessibility of the strong coupling
\citep{jin_controlling_2003}. A typical nanoprism has dimensions
of about 60-80 nm in side length and about 10 nm in thickness. Such
nanoprism dimensions result in a plasmon resonance, $\omega_{pl}$,
overlapping with the exciton transition, $\omega_{X^{0}}$, in the
monolayer WS\textsubscript{2} in the coupled system. Thus by combining
these two materials one could expect strong plasmon-exciton interactions.
In order to verify this expectation, we positioned Ag nanoprisms on
top of the monolayer by drop casting a nanoparticle suspension on
a polymer pre-coated substrate (see Methods). As a result we obtain
a WS\textsubscript{2} flake covered with Ag nanoprisms of various
sizes as is evidenced by the dark-field (DF) optical microscopy (see
Fig. \ref{fig:1}c). The colourful spots in the image are individual
Ag nanoprisms possessing different plasmon resonances. This allows
us to study a variety of plasmon-exciton resonance detunings $\delta=\omega_{pl}-\omega_{X^{0}}$
within the same sample. A combined 2D material - silver nanoprism
system is depicted schematically in Fig. \ref{fig:1}d. The optical
states are localized at the corners of the prism and are shown schematically
by the bright spots. Insets show a magnified DF image of a single
nanoprism and an SEM image of the coupled system. The latter confirms
that an individual nanoprism is measured in the optical microscope.

\begin{figure}[H]
\hspace{20mm}\includegraphics[width=120mm]{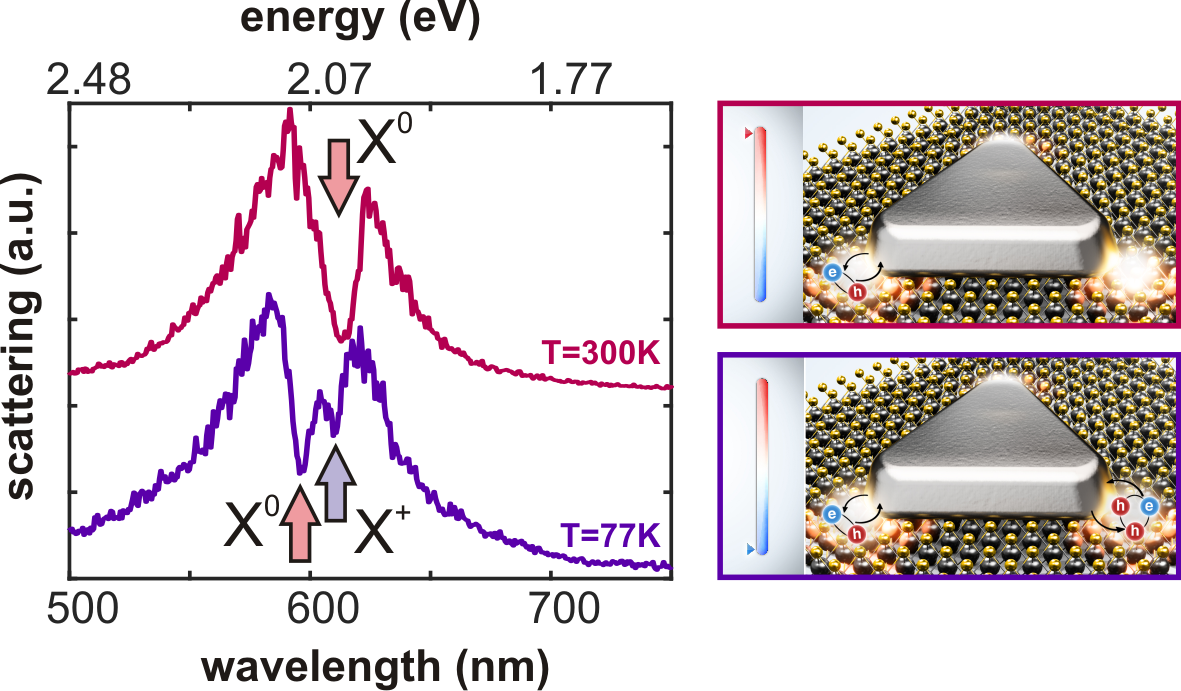}

\caption{\textbf{\footnotesize{}Coupled system under room and liquid nitrogen
temperature.}{\footnotesize{} Left: Dark-field scattering spectra
at 77 and 300 K. At ambient conditions the DF spectrum depicts two
peaks, namely the upper and lower polaritons. At liquid nitrogen temperature
the DF spectrum shows three peaks that are identified as upper middle
and lower polaritons. Arrows show neutral (}$X^{0}${\footnotesize{})
and charged (}$X^{+}${\footnotesize{}) exciton resonance wavelengths.
Right: Artist view of plasmon-exciton mixture at T=300 K and plasmon-exciton-trion
interaction at T=77 K.\label{fig:2}}}
\end{figure}

\textbf{Temperature dependence.} We now explore the evolution of the
strongly coupled system by varying its temperature (see Methods for
experimental details). Since the PL spectrum exhibits the appearance
of a trion resonance at low temperature, one could expect strong interaction
between the trion and the plasmon. The DF spectrum of the hybrid system
(Fig. \ref{fig:2}) measured at ambient conditions reveals the splitting
of the resonance into two peaks, which demonstrates the formation
of plasmon-exciton polaritons. The high-energy peak represents the
upper polariton and the low-energy peak is the lower polariton. The
dip between the polaritonic branches coincides with $\omega_{X^{0}}$.
Upon decreasing the temperature down to liquid nitrogen (77 K), the
charged state of the exciton is clearly observed and starts dominating
the PL spectrum (see Fig. \ref{fig:1}b and SI Fig. S3). As this resonance
is split from the exciton by only $\sim$40 meV, the strong interaction
of both exciton and trion resonances with the plasmonic cavity is
plausible. This is revealed by the appearance of two dips in the DF
spectrum that match the energies of the exciton and trion states (see
Fig. \ref{fig:2} T=77 K). 

The hybrid system thus experiences a transition from plasmon-exciton
interaction at room temperature to a more complex plasmon-exciton-trion
interaction at low temperature, which is schematically illustrated
in Fig. \ref{fig:2} (right panel). This evolution was studied by
measuring the DF scattering as a function of temperature. The results
for the hybrid system with approximately zero detuning (the detuning
is defined at 77 K) are presented in Fig. 3a. The cases of positive
and negative detuning are shown in the SI (Fig. S2). Irrespectively
of the detuning, however, the temperature induced modifications are
similar: two dips are clearly observed at low temperature, which demonstrate
that both the neutral exciton (the high-energy dip) and the trion
(the low-energy dip) interact strongly with the plasmonic cavity.

The coupled plasmon-exciton-trion system thus results in a formation
of three bright hybrid states \textendash{} upper, middle and lower
polaritons. The corresponding exciton, trion and plasmon contributions
can in principle be inferred from the spectral data. Notably, all
polaritons possess non-vanishing contributions of all three system
subparts \textendash{} excitonic, trionic and plasmonic. For this
reason these states are potentially interesting for charge transport
as they carry non-zero net electrical charge (positive in this case).
The composition of these states can be understood by diagonalyzing
the 3\texttimes 3 Hamiltonian of the system \citep{rapaport_negatively_2000}:
\begin{equation}
\left(\begin{array}{ccc}
\omega_{pl}-i\gamma_{pl} & g_{X^{0}} & g_{X^{+}}\\
g_{X^{0}} & \omega_{X^{0}}-i\gamma_{X^{0}} & 0\\
g_{X^{+}} & 0 & \omega_{X^{+}}-i\gamma_{X^{+}}
\end{array}\right)
\end{equation}
\\
with $g_{X^{0}}$ and $g_{X^{+}}$ being the interaction constants
of plasmon with exciton and trion, respectively and $\gamma_{X^{0}}$
$(\gamma_{X^{+}})$ corresponds to the dissipation rate of the exciton
(trion). The main conclusion of this analysis is that for the appropriate
parameters there exists a solution with three bright polariton states.
The middle polariton has a relatively small, \ensuremath{\sim}20\%,
plasmon component (which vanishes completely in the very strong coupling
regime), while both upper and lower polaritons are composed roughly
of \ensuremath{\sim}50\% plasmon and of \ensuremath{\sim}25\% exciton
and \ensuremath{\sim}25\% trion respectively (see SI, Fig. S7).

As the temperature rises, the dip associated with the trion state
becomes less pronounced and completely disappears at 300 K. Such behavior
is consistent with the PL measurements and reflectivity data of the
bare monolayer WS\textsubscript{2} (see SI Fig. S3) and can be explained
by taking into account the loss of oscillator strength by the trion
as the temperature is increased \citep{arora_excitonic_2015}. Additionally,
note that the binding (dissociation) energy of the trion is about
40 meV, which is comparable to k\textsubscript{B}T at room temperature
thus making the trion less stable at elevated temperatures.

\begin{figure}[H]
\hspace{20mm}\includegraphics[width=120mm]{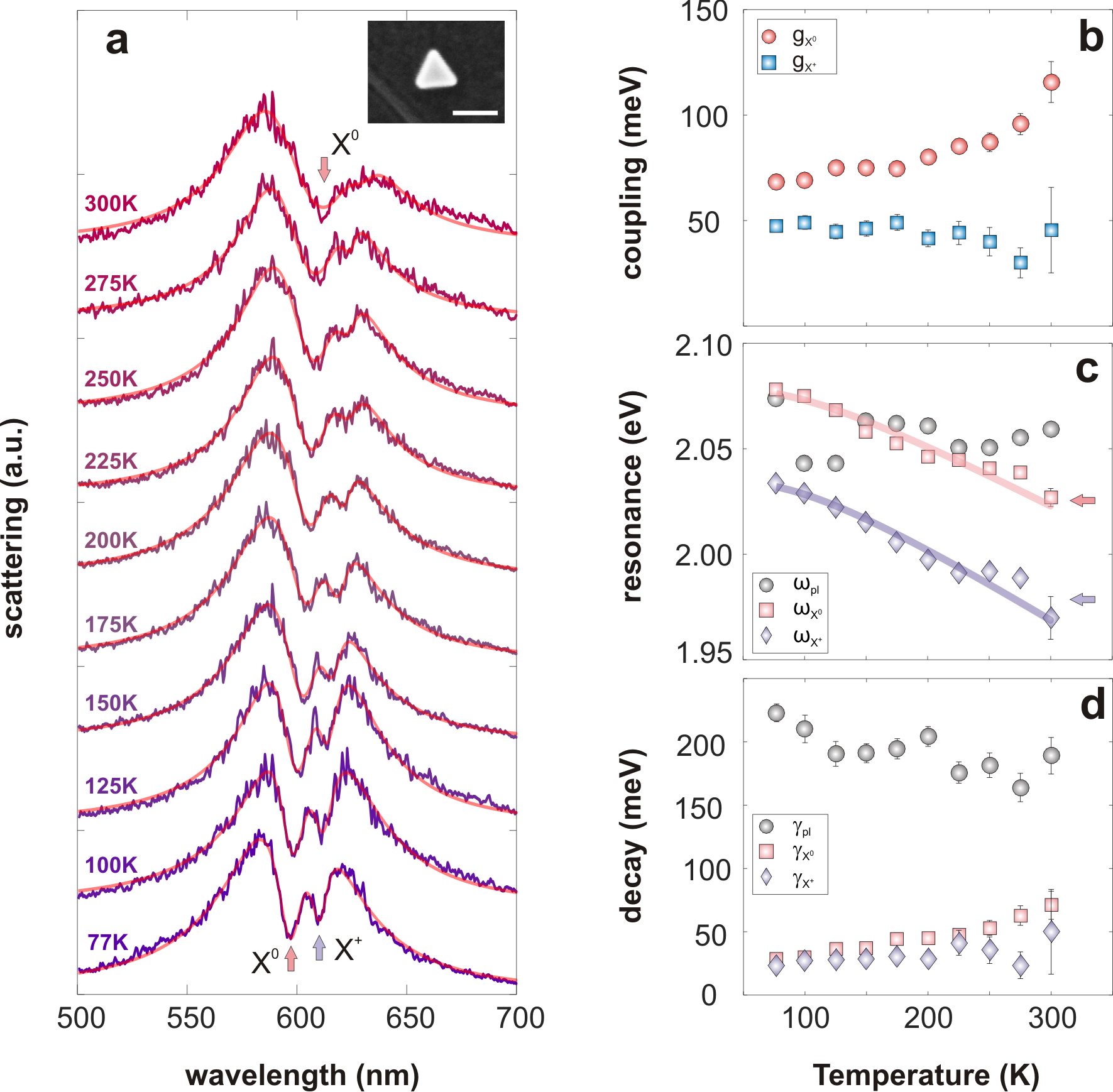}

\caption{\textbf{\footnotesize{}Temperature dependence of the coupled system.}{\footnotesize{}
(a) Dark field spectra for an individual silver nanoprism \textendash{}
monolayer WS\protect\textsubscript{{\footnotesize{}2}} hybrid as
a function of temperature. Semi-transparent red curves show coupled
oscillator model fits of the data. Inset: SEM image of the corresponding
nanostructure (scale bar - 100 nm). (b-d) Temperature dependence of
the coupled oscillator model parameters. (b) Coupling rates for exciton
and trion, (c) resonance energies of plasmon, exciton and trion resonances.
Solid lines show a fit using the semiconductor model. The O\textquoteright Donnel
model gives the following values of $E_{g}(0)=$2.08 eV (2.03 eV),
S=1.82 (2.17) and $\left\langle \hbar\omega\right\rangle =$26 meV
(26 meV) for }$X^{0}${\footnotesize{} (}$X^{+}${\footnotesize{})
correspondingly for the coupled case. (d) Decay rates for plasmon,
exciton and trion correspondingly.\label{fig:3}}}
\end{figure}

The strong interaction between the plasmon and neutral exciton, however,
remains throughout the whole temperature range studied. This is likely
due to the fact that the exciton increases its oscillator strength
as the temperature rises. It is also important to note that the binding
energy of the exciton, which is on the order of \ensuremath{\sim}700
meV, is much above the thermal energy and thus thermal dissociation
and broadening does not prevent the coupling even at room temperature.

The temperature dependence of the oscillator strength can be understood
by noting that the oscillator strength depends on the number of available
states. By decreasing the temperature the distribution of states changes
in favour of trions. Thus the trions gain and the neutral excitons
loose the oscillator strength upon cooling \citep{arora_excitonic_2015}.
This is consistent with our observations.

\textbf{Coupled oscillator model.} We now turn to an in-depth analysis
of the temperature scans. We use a modified coupled harmonic oscillator
model to extract the parameters of the hybrid system. This model considers
a cavity driven by an external field, which is coupled to two dissipative
oscillators \citep{wu_quantum-dot-induced_2010,zengin_approaching_2013}
(see Methods). In this case, we consider coupling to neutral and charged
excitons. The scattering cross-section of such a system is then given
by:

\begin{equation}
\sigma_{scat}\propto\left|\frac{\omega^{2}\widetilde{\omega}_{X^{0}}^{2}\widetilde{\omega}_{X^{+}}^{2}}{\widetilde{\omega}_{X^{0}}^{2}\widetilde{\omega}_{pl}^{2}\widetilde{\omega}_{X^{+}}^{2}-\omega^{2}\left(g_{X^{0}}^{2}\widetilde{\omega}_{X^{+}}^{2}+g_{X^{+}}^{2}\widetilde{\omega}_{X^{0}}^{2}\right)}\right|^{2}
\end{equation}

with

\[
\begin{array}{c}
\widetilde{\omega}_{X^{0}}^{2}=\left(\omega^{2}-\omega_{X^{0}}^{2}+i\gamma_{X^{0}}\omega\right),\\
\begin{array}{c}
\widetilde{\omega}_{X^{+}}^{2}=\left(\omega^{2}-\omega_{X^{+}}^{2}+i\gamma_{X^{+}}\omega\right),\end{array}\\
\widetilde{\omega}_{pl}^{2}=\left(\omega^{2}-\omega_{pl}^{2}+i\gamma_{pl}\omega\right)
\end{array}
\]
\\
being the harmonic oscillator terms for the exciton, trion and plasmon,
respectively.

We further utilize Eq. (2) to fit the temperature-dependent DF scattering
spectra. We observe that the coupled oscillator model reproduces the
experimental results exceptionally well (see semi-transparent red
curves in Fig. \ref{fig:3}a). The parameters extracted from the coupled
oscillator model as a function of temperature are shown in Fig. \ref{fig:3}b-d.

The first important observation is that the coupling to the neutral
excitons $g_{X^{0}}$ drops, whereas the coupling to the trions $g_{X^{+}}$
increases upon cooling (Fig. \ref{fig:3}b). This ultimately leads
to the emergence of the two dips in the scattering spectra at low
temperature. At the same time at high temperature, coupling to trions
essentially vanishes. This is confirmed by a substantial degradation
of the fit quality for T=300 K (note the error bars for the trion
coupling rate at T=300 K). This is in agreement with the qualitative
analysis above, where we argued that the trion becomes less stable
at room temperature. Notably, at room temperature the interaction
becomes rather weak and Fano-like despite the relatively high value
of the coupling strength with respect to decoherence rates of the
plasmon and the exciton. This is because of the large plasmon-exciton
detuning at room temperature (see Fig. \ref{fig:3}a,c T=300 K).

Another important observation is that both the trion and the exciton
resonances blue-shift upon cooling (Fig. \ref{fig:3}c). This is in
agreement with the standard semiconductor behavior \citep{oddonell_temperature_1991}
and is confirmed here on the bare monolayer WS\textsubscript{2} (see
SI Fig. S3 and S5). The fact that the exciton and trion resonances
follow the temperature dependence that is consistent with the standard
semiconductor behavior even in the coupled system signals that the
coupled oscillator model provides a meaningful physical picture and
therefore justifies its use. Indeed, temperature dependence of the
semiconductor bandgap is conveniently described by the O\textquoteright Donell
model for the temperature dependency of the semiconductor energy gap
\citep{oddonell_temperature_1991},

\begin{equation}
E_{g}\left(T\right)=E_{g}\left(0\right)-S\left\langle \hbar\omega\right\rangle \left\{ coth\left[\nicefrac{\left\langle \hbar\omega\right\rangle }{2k_{B}T}\right]-1\right\} 
\end{equation}

where $E_{g}\left(0\right)$ is the bandgap at 0 K, $S$ is the electron-phonon
coupling strength, and $\left\langle \hbar\omega\right\rangle $ is
the average phonon energy. This model fits our data both coupled (Fig.
\ref{fig:3}c) and uncoupled (see SI, Fig. S5) very well. The obtained
values are consistent with earlier studies \citep{ross_electrical_2013},
which again justifies the use of the coupled oscillator model. Furthermore,
we observe that the dissociation energy, $\omega_{X^{0}}-\omega_{X^{+}}$,
is nearly constant in the studied temperature range and is about 40
meV (see Fig. 3c and Fig. S3).

We note that for the uncoupled nanoparticle system the plasmon resonance
is expected to red shift upon cooling \citep{liu_reduced_2009} (see
SI, Fig. S1). However, in the coupled system this behavior is not
observed (see Fig. \ref{fig:3}c). We argue that this is due to interaction
with the TMDC material, which \textquotedblleft pulls\textquotedblright{}
the plasmon resonance to higher energies at low temperature. In other
words, exciton and trion resonances follow the standard semiconductor
temperature dependence in accordance with Eq. (2), while the plasmon
passively follows the change in its immediate dielectric environment
induced by temperature changes.

\begin{figure}[H]
\hspace{20mm}\includegraphics[width=120mm]{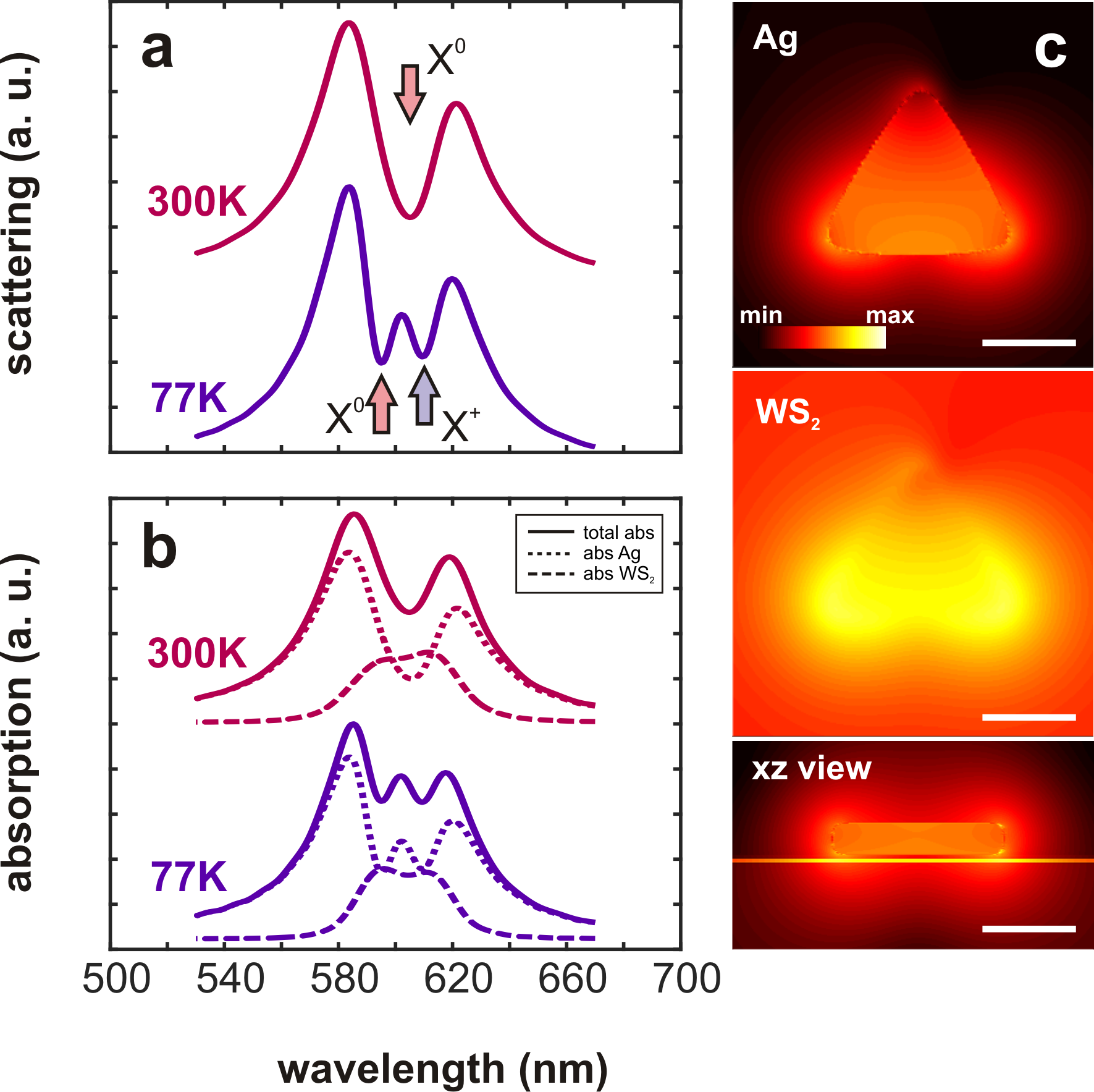}

\caption{\textbf{\footnotesize{}Numerical simulation.}{\footnotesize{} (a)
Scattering cross-sections spectra as a function of temperature. At
T=300 K the dielectric function of WS\protect\textsubscript{{\footnotesize{}2}}
was assumed to have only one resonance energy corresponding to uncharged
exciton }$X^{0}${\footnotesize{}, while at T=77 K both }$X^{+}${\footnotesize{}
and }$X^{0}${\footnotesize{} contribute to the signal. Parameters
of the model: $\varepsilon_{Ag}(\omega)$ dielectric function for
silver is taken from Palik \citep{palik}, WS\protect\textsubscript{{\footnotesize{}2}}
from Ref. \citep{li_measurement_2014}, nanoprism side length L=70
nm, height H=10 nm, nanoprism edges are rounded with radius of curvature
$r\backsim$ 5 nm. TMDC layer was positioned on a dielectric slab
with refracting index $n=$1.5 that mimics a glass substrate. The
whole system is embedded in vacuum. (b) Total, silver and WS\protect\textsubscript{{\footnotesize{}2}}
absorption cross-sections spectra as a function of temperature. Note
splitting observed at both temperatures, which confirms the intermediate
coupling regime. (c) Electromagnetic energy density distribution,
$\rho(r)$, within the nanoparticle and the 2D material at $\lambda=$
577 nm. Both top and side views are shown. Scale bar = 30 nm.}\label{fig:4} }
\end{figure}

We further extract and compare the decoherence rates to elucidate
on whether or not the strong coupling condition is met (see Figs.
\ref{fig:3}b,d). We note that the strict condition for the strong
coupling regime, that is, $2g_{X^{0}}>\gamma_{pl}$, is satisfied
for neutral excitons but not for trions $(2g_{X^{+}}>\gamma_{pl}$)
at high temperature. As we show in the SI this regime manifests itself
through several observable Rabi cycles (see Fig. S6). At high temperatures
the trion is in the weak coupling regime as the coupling becomes less
than both cavity and trion dissipation rates $2g_{X^{+}}<\gamma_{pl},\gamma_{X^{+}}$
\citep{dufferwiel_excitonpolaritons_2015}. At high temperatures,
however, the system is in the intermediate regime for the independent
exciton and trion contributions \textendash{} $\gamma_{pl}>2g_{X^{0}}>\gamma_{X^{0}}$
(or $\gamma_{pl}>2g_{X^{+}}>\gamma_{X^{+}}$) \citep{zengin_approaching_2013,antosiewicz_plasmonexciton_2014},
while the combined coupling still exceeds the strong coupling limit,
i.e. $2\sqrt{g_{X^{0}}^{2}+g_{X^{+}}^{2}}\gtrsim\gamma_{pl}$. We
thus conclude that the combined plasmon-exciton-trion system reaches
the strong coupling regime. Noteworthy, the plasmon-trion coupling
here, $g_{X^{+}}\thickapprox$50 meV, exceeds the trion binding energy,
$\omega_{X^{0}}-\omega_{X^{+}}\thickapprox$40 meV, which is an interesting
regime since the light-matter coupling surpasses the Coulomb interaction
in this case. It was recently predicted that in this regime the trion
orbital and spin properties are strongly modified \citep{grenier_trion-polaritons_2015}.
In the following discussion, we calculate the scattering and absorption
cross-sections of the coupled system to demonstrate this point more
rigorously (Fig. \ref{fig:4}).

In further analysis (Fig. \ref{fig:3}d), we notice that the linewidth
of excitons and trions reduces upon cooling, thus making the corresponding
scattering dips more pronounced. Exciton line narrowing upon cooling
is a well-documented behavior consistent with previous results \citep{moody_intrinsic_2015,dey_optical_2016}
and confirmed here for the uncoupled exciton and trion system (see
SI, Fig. S3). At the same time plasmons in the coupled system do not
narrow upon cooling (Fig. \ref{fig:3}d). This is somewhat surprising
considering that uncoupled plasmons are anticipated to red shift and
slightly narrow upon cooling as was reported previously \citep{liu_reduced_2009}
and confirmed here in a control experiment (see SI Fig. S1). Such
behavior is likely caused by the weak interaction with the TMDC material.
Indeed both plasmon resonance and its full width half maximum depend
on the electromagnetic environment, which is temperature-dependent
in this case. 

In closing the analysis section, we note that Fig. \ref{fig:3} corresponds
to the case of zero detuning. Similar data is presented in the SI
for the case of positive and negative detuning (see Fig. S2). These
observations are reproducible and consistent with the coupled oscillator
model. The data shows increase of trion coupling while reduction of
exciton coupling upon cooling. At the same time, both the exciton
and trion resonances narrow upon cooling. The data shows consistent
blue shift in the exciton and the trion resonance upon cooling in
agreement with the standard semiconductor behavior. Most notably,
however, the data demonstrates reproducible coupling to the charged
exciton at low temperature. This latter fact makes us confident to
conclude we observe formation of charged polaritons in this system.

\textbf{Numerical calculations.} To gain further insight into the
nature of plasmon-exciton-trion interactions in this system, we perform
numerical calculations using the finite-difference time domain (FDTD)
method (see Methods). The results reproduce all essential features
of the experiment and the analytical coupled oscillator model (see
Fig. \ref{fig:4}). This implies that, despite significant simplifications
used in the coupled oscillator model, it adequately describes the
data, even in a rather complex regime of two types of excitons involved
in the coupling process such as studied here. We thus emphasize the
power of the coupled oscillator model in describing such system.

Comparison between experimental observations and the FDTD calculations
allows to estimate the oscillator strength corresponding to both neutral
and charged excitons as well as to compare them to the values obtained
from reflectivity and photoluminescence (see Fig. S3). The dielectric
susceptibility of WS\textsubscript{2} monolayer at room temperature
(corresponding solely due to neutral exciton contribution) was extracted
from the literature values and was used here without any further open
parameters \citep{li_measurement_2014}. The agreement between the
FDTD calculations and experimental observations in Figs. \ref{fig:2}-\ref{fig:4}
is striking. The low temperature $\varepsilon(\omega)$ of WS\textsubscript{2}
was introduced phenomenologically (by two Lorentzian contributions),
yet it also leads to a good agreement between the experiment and the
theory. Thus the phenomenological $\varepsilon(\omega)$ at low temperature
corresponds to the coupling rates of both neutral and charged excitons
extracted from the coupled oscillator model (Fig. \ref{fig:3}b).

In addition to the calculated scattering spectra (Fig. \ref{fig:4}a),
we perform calculations of absorption cross-section of the coupled
system as well as its components \textendash{} Ag nanoprism and WS\textsubscript{2}
monolayer (Fig. \ref{fig:4}b). The plots in Fig. \ref{fig:4}b clearly
show splitting of absorption in the Ag nanoprism, as well as smaller
but visible splitting in the absorption spectra of the WS\textsubscript{2}
monolayer. This is a signature of strong coupling \citep{antosiewicz_plasmonexciton_2014}.
The small splitting in the 2D material absorption is explained by
a lack of long-lived oscillations between Ag and WS\textsubscript{2}.
As shown in Fig. S6, the time evolution of the Rabi oscillations is
short-lived and exhibits only one significant period. A similar observation
can be drawn from the oscillations of the electric fields around the
Ag triangle, where the excitation energy alternates between the plasmon-polariton
(visible enhanced fields) and the 2D material (see SI movies).

We further inspect the electromagnetic energy density distributions,
{\footnotesize{}$\rho(r)$,} \citep{ruppin_electromagnetic_2002}
in the coupled system (Fig. \ref{fig:4}c). We observe that the mode
in the nanoprism is confined mainly to its interior. We also observe
that while the energy density in the WS\textsubscript{2} is greater
than in Ag, the monolayer occupies only a small volume due to its
two-dimensionality and therefore the total energy stored in the metal
and the 2D material is similar. For single nanoparticles the mode
is concentrated within its volume \citep{koenderink_use_2010,sauvan_theory_2013,kristensen_modes_2014}
and to obtain long-lived Rabi oscillations the excitonic material
needs not only a large transition dipole moment, but also needs to
saturate the small fraction of the mode present outside the metal
resonator \citep{yang_role_2016,zengin_realizing_2015}. In the present
case this is not possible (Fig. \ref{fig:4}c), yet filling up a small
fraction of the mode volume with the 2D material is enough to reach
the strong coupling. This is likely due to the large transition dipole
moment of WS\textsubscript{2} ($\mu_{e}=$56 Debye \citep{sie_valley-selective_2015}).
We thus conclude that a coupled plasmon-exciton polariton system comprising
a silver nanoprism and monolayer WS\textsubscript{2} is very useful
for practical realization of strong coupling. One may ask how many
excitons contribute to the coupling process in this case, especially
taking in mind the large transition dipole moment of WS\textsubscript{2}.
Following the well-known relation for the coupling strength $g=\sqrt{N}\mu_{e}|E_{vac}|=\sqrt{N}\mu_{e}\sqrt{\hbar\omega/(2\varepsilon\varepsilon_{0}V)}$
and our previous estimations using J-aggregates \citep{zengin_realizing_2015},
we obtain $N\sim10-20$ for realistic mode volumes in the range of
$V=(2-4)\times$$10^{4}$nm\textsuperscript{3}. This number closely
approaches the quantum optics limit thus making this system potentially
interesting for photon-photon interactions.
\\
\\
\textbf{DISCUSSION}
\\
To conclude, we have demonstrated that the optical response of the
strongly coupled system can be tailored by tuning the temperature.
At low temperatures the plasmonic cavity strongly interacts with both
neutral and charged excitons in the monolayer WS\textsubscript{2},
whereas at room temperature only the neutral exciton is coupled to
the plasmon. The hybrid structure in this study consists of an individual
silver nanoprism and two-dimensional monolayer WS\textsubscript{2}.
This system is extremely compact, yet able to strongly interact with
light.

We note that the temperature-dependent PL signal of the WS\textsubscript{2}
monolayer and in particular the second resonance at low energy appearing
upon decreasing of temperature follows the trend associated with the
trion state \citep{ross_electrical_2013}. Other low energy states
like bound states or localized excitons \citep{jones_optical_2013,wang_valley_2014}
have lower dissociation energies and are not stable at elevated temperatures
\citep{ganchev_three-particle_2015}. Based on this, we conclude that
the low energy peak observed in the PL spectrum (Fig. \ref{fig:1}b)
and the low energy dip in the scattering spectra (Figs. \ref{fig:2}-\ref{fig:3})
indeed appear due to the positively charged exciton \textendash{}
i.e. the trion. Here, the trion state is stabilized by low temperature
and chemical doping \citep{mouri_tunable_2013}.

Our findings demonstrate the principal possibility of studying electrically
charged polaritons in a form of plasmon-exciton-trion hybrids. Here,
this is done at low temperature to ensure the stability of the charged
exciton state, however, by stabilization of the trion state by other
means, for example by electrostatic doping, similar effects can be
anticipated at room temperature \citep{zhu_exciton_2015}. Constructing
macroscopic coherent polariton states that will constitute a coherent
mixture of excitons, trions and cavity excitations may find use in
charge transport and optoelectronic devices by boosting the carrier
mobility \citep{orgiu_conductivity_2015} in analogous way to the
theoretically predicted exciton transport enhancement mediated by
a cavity \citep{feist_extraordinary_2015,schachenmayer_cavity-enhanced_2015}.
This new degree of freedom \textendash{} charged polaritons is the
central observation of this work. We envision that these findings
may find potential implications for various optoelectronic applications,
such as light harvesting and light emitting devices.
\\
\\
\textbf{METHODS}
\\
\textbf{Sample preparation:} Silver nanoprisms were synthesized using
a seed-mediated protocol \citep{jin_controlling_2003}. WS\textsubscript{2}
were mechanically exfoliated from bulk crystal (HQ-graphene) and transferred
to a Si wafer with a thermally oxide layer (50 nm-thick) using the
all-dry transfer method \citep{castellanos_all_dry}. Monolayers were
identified by photoluminescence spectra and optical contrast. In order
to ensure a proper density of nanoprisms covering the monolayer flake,
an adhesion layer was deposited before the nanoprism (Polylisine 0.25
mg/mL), then nanoparticles are drop-cast and let them rest for 2 min,
the excess of solution is gently removed by an absorptive tissue.
We note that the polymer layer that we utilize to adhere silver nanoprisms
to the monolayer WS\textsubscript{2} plays an important role in the
stabilization of the trions. This is likely induced by the chemical
doping \citep{mouri_tunable_2013}. The trion in this case is mostly
a positively charged state since the adhesion layer tends to p-dope
WS\textsubscript{2}. To demonstrate this, in the SI (see Fig. S5)
we perform a control experiment which shows that strong plasmon-trion
coupling is not observed in the absence of the poly-lysine adhesion
layer.

\textbf{Optical measurements:} Dark-field optical microscopy and spectroscopy
measurements were done using a laser driven light source (ENERGETIQ
EQ-99XFC) with side illumination configuration at an angle of about
50 degrees. For photoluminescence experiments, the sample was excited
by a CW 532 nm (2.33 eV) laser under irradiance of \ensuremath{\sim}100
W/cm\textsuperscript{2}. PL and DF signals were collected using a
20\texttimes{} objective (Nikon, NA=0.45) and directed to a fibre-coupled
30 cm spectrometer (Andor Shamrock SR-303i) equipped with a CCD detector
(Andor iDus 420). Low temperature measurements were performed using
a cold finger optical cryostat (Janis).

\textbf{Coupled oscillators model:} Each component (plasmonic nanoparticle,
exciton, and trion) of the coupled system is described as a harmonic
oscillator with its own resonance frequency and damping. Coupling
between the nanoparticle and WS\textsubscript{2} excitations is mediated
via inductive terms proportional to the oscillator velocities. Dynamics
of the full system is therefore modeled by three coupled equations
\citep{wu_quantum-dot-induced_2010}:

\[
\begin{array}{c}
\ddot{x}_{pl}+\gamma_{pl}\dot{x}_{pl}+\omega_{pl}^{2}x_{pl}+g_{X^{0}}\dot{x}_{X^{0}}+g_{X^{+}}\dot{x}_{X^{+}}=-eE(t)\\
\ddot{x}_{X^{0}}+\gamma_{X^{0}}\dot{x}_{X^{0}}+\omega_{X^{0}}^{2}x_{X^{0}}-g{}_{X^{0}}\dot{x}_{pl}=0\\
\ddot{x}_{X^{+}}+\gamma_{X^{+}}\dot{x}_{X^{+}}+\omega_{X^{+}}^{2}x_{X^{+}}-g{}_{X^{+}}\dot{x}_{pl}=0
\end{array}
\]
\\

Here, $x_{pl}$, $x_{X^{0}}$ , and $x_{X^{+}}$ represent coordinates
of the cavity, exciton, and trion oscillators, respectively, $E(t)$
is the driving electric field, and $e$ is the elementary charge.
We assume that only nanoantenna interacts directly with the incident
field which reflects the fact that exciton and trion extinction is
negligible compared to that of the plasmonic nanostructure \citep{wu_quantum-dot-induced_2010}.
The scattering cross-section of the coupled system, being dominated
by the nanoparticle dipole moment radiation, can be estimated in this
phenomenological approach as $\sigma_{scat}\propto\left|e\ddot{x}_{pl}\right|^{2}\propto\omega^{4}\left|x_{pl}\right|^{2}$.
Finally, we note that this approach can be easily generalized to an
arbitrary number of coupled oscillators.
\\
\\
\textbf{REFERNCES}

\bibliographystyle{apsrev4-1}
\bibliography{ref_A}

\newpage

\renewcommand{\theequation}{S\arabic{equation}}
\renewcommand{\thefigure}{S\arabic{figure}}
\setcounter{figure}{0}
\setcounter{equation}{0}

\begin{widetext}

{\centering \large{\textbf{Supplemental Information for: \\Observation of tunable charged exciton polaritons in hybrid monolayer
WS\textsubscript{2} \textendash{} plasmonic nanoantenna system}}

\normalsize{Jorge Cuadra, Denis G. Baranov, Martin Wers\"all, Ruggero Verre, Tomasz J. Antosiewicz, and Timur Shegai

\emph{Department of Applied Physics, Chalmers University of Technology, 412 06 G\"oteborg, Sweden}

\emph{Centre of New Technologies, University of Warsaw, Banacha 2c, 02-097 Warszawa, Poland}}

}

\end{widetext}

\tableofcontents

\clearpage

\subsubsection{~~Bare nanoprism data:}

Figure S1 displays the DF spectra as a function of temperature for
an individual bare Ag nanoprism. The decrease of the temperature from
300 K until about 225 K (Debye temperature T\textsubscript{\textgreek{j}}
for bulk silver is \ensuremath{\sim}235 K) causes a red-shift of the
plasmonic resonance. Below T\textsubscript{\textgreek{j}}, damping
due to electron-phonon scattering is reduced and the plasmon\textquoteright s
resonance barely moves and exhibits a saturation \citesupp{liu_reduced_2009_s}. 

\begin{figure}[H]
\centering
\includegraphics[width=120mm]{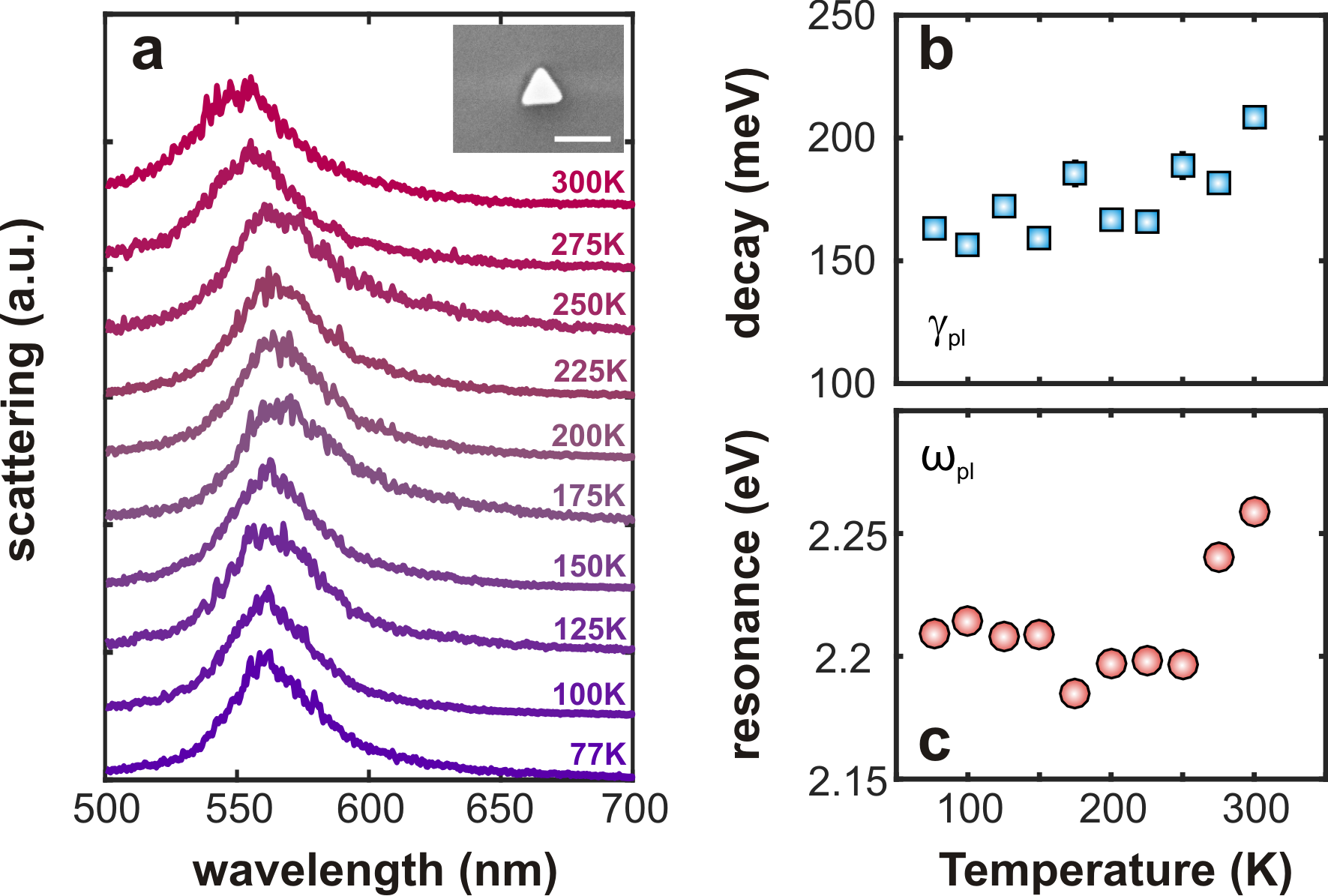}

 \caption{Temperature dependency of the uncoupled systems. a) DF spectra as
a function of temperature for a silver prism on Si/SiO\protect\textsubscript{2}substrate.
The plasmonic resonance shifts its position towards low energies as
the temperature is decreased. Inset corresponds to the SEM image showing
the nanoprism; scale bar 50 nm. b) Temperature dependence of the PL
for the same flake shown in Fig. 2. At 300 K the spectrum is dominated
by the exciton whereas the trion is weaker. As the temperature is
decreased both resonances blue-shift. At 77 K the trion dominates
the PL spectrum.}
\end{figure}

\clearpage{}

\subsubsection{~~DF for positive and negative detuning:}

\begin{figure}[H]
\centering
\includegraphics[width=15cm]{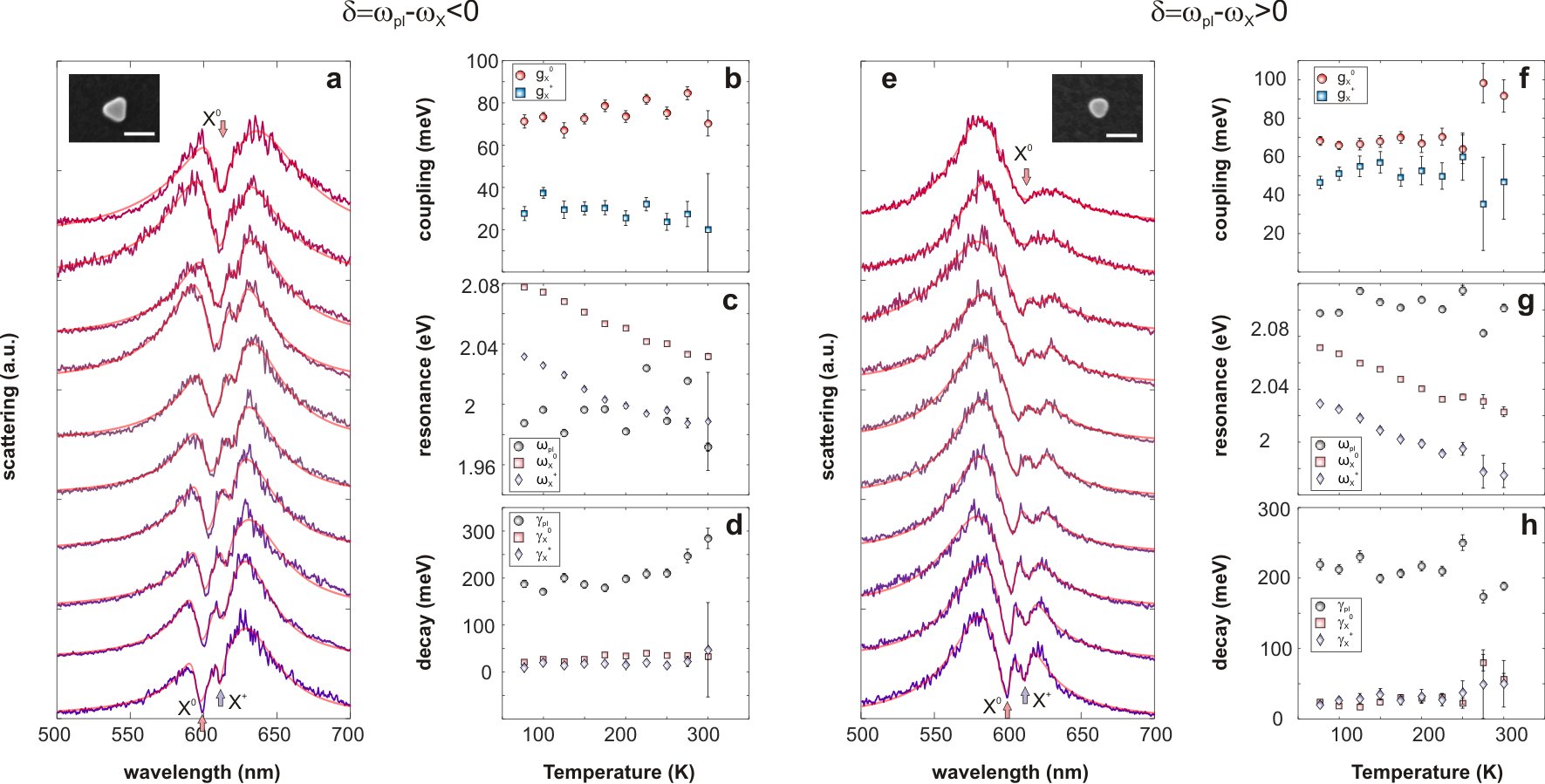}

\caption{Temperature dependence of the coupled system for positive and negative
detuning. (a-d) the case of negative detuning, (e-h) the case of positive
detuning. (a, e) Dark field spectra for an individual silver nanoprism
\textendash{} monolayer WS\protect\textsubscript{2} hybrid as a function
of temperature. Semi-transparent red curves show coupled oscillator
model fits of the data. Inset: SEM image of the corresponding nanostructure
(scale bar - 100 nm). (b-d and f-h) Temperature dependence of the
coupled oscillator model parameters. (b, f) coupling rates for exciton
and trion, (c, g) resonance energies of plasmon, exciton and trion
resonances, (d, h) decay rates for plasmon, exciton and trion correspondingly.}
\end{figure}

\clearpage{}

\subsubsection{~~Reflectivity and photoluminescence of monolayer WS2 vs temperature
and vs polymer adhesion layer:}

\begin{figure}[H]
\centering
\includegraphics[width=80mm]{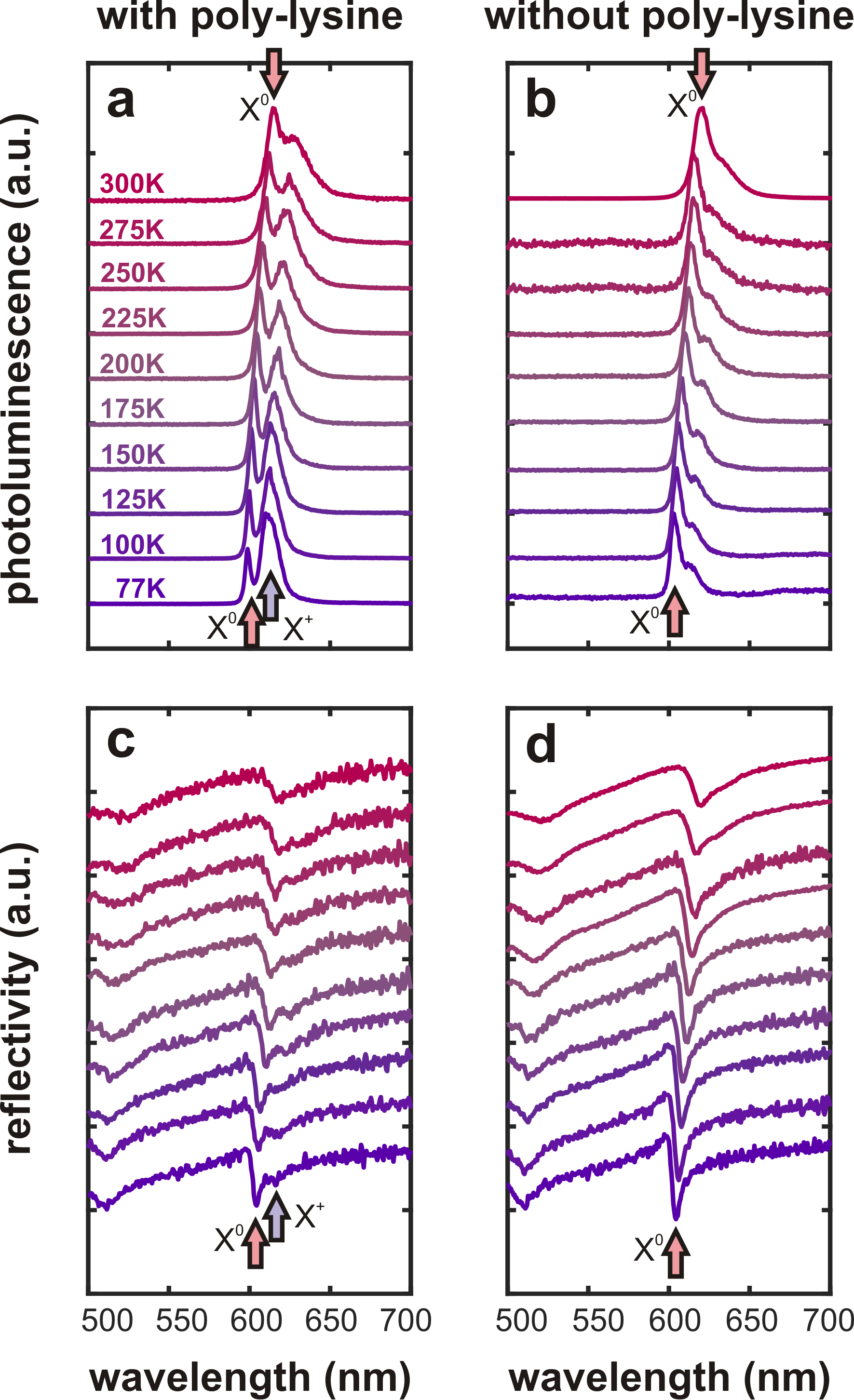}

\caption{Reflectivity measurements. (a) PL spectra for T = 77 - 300 K when
the polymer adhesion layer was used. At low temperature both exciton
and trion resonances are well resolved. As the temperature is raised
the trion is less visible. (b) Same as in (a) but without the polymer
adhesion layer. The trion resonance is not observed for any temperature,
whereas the exciton is observed for all temperatures as in a). (c)
Reflectivity spectra for T = 77 - 300 K when the polymer adhesion
layer is used. A- and B-excitons are seen for all temperatures. For
the A-exciton at low temperature both exciton and trion resonances
are resolved, as the temperature is raised the trion is less visible.
(d) Same as in (c) but without the polymer adhesion layer. The trion
resonance is not observed in reflectivity measurements for any temperature,
whereas the exciton is observed for all temperatures as in c).}
\end{figure}

Role of the polymer adhesion layer: The polymer adhesion layer not
only ensures a proper density of nanoprisms but also changes the excitonic
specimens (Fig. S3). For samples with adhesion layer the PL at low
temperature is dominated by the trion whereas in samples without adhesion
layer the exciton prevails in the PL spectra. This is caused by a
chemical doping of p-type carriers (positively charged poly-lysine
molecules) and stabilization of a positively charged trion state in
the system \citesupp{mouri_tunable_2013_s}.

\clearpage{}

\subsubsection{~~Coupled system without polymer adhesion layer}

\begin{figure}[H]
\centering
\includegraphics[width=78.5mm]{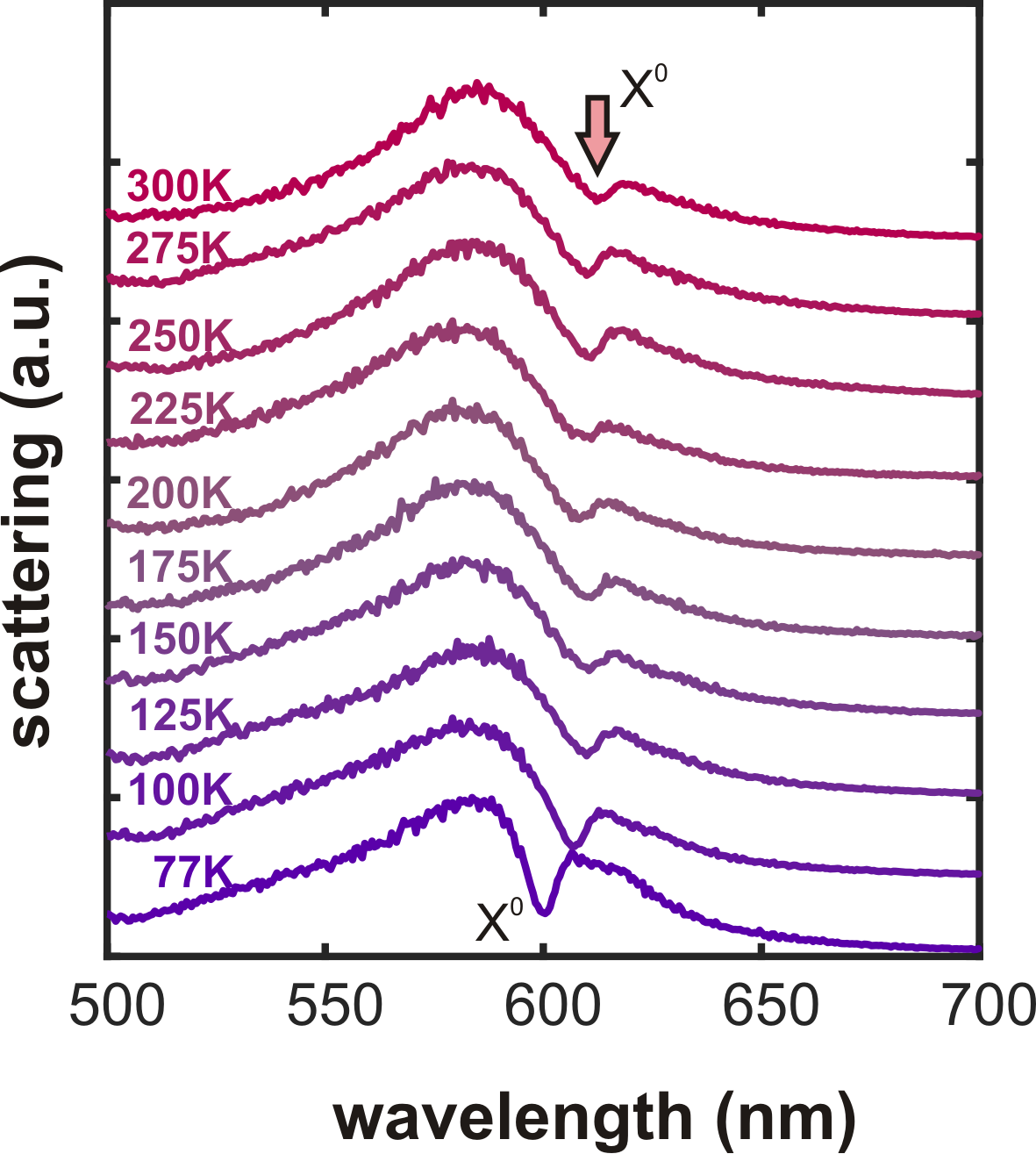}
\caption{Samples without polymer adhesion layer. (a) Dark field spectra as
a function of the temperature shows the interaction with the neutral
exciton state. (b) PL spectra for the uncoupled monolayer without
adhesion layer. Inset in a) depicts the SEM of the prism on top of
the monolayer; scale bar 50 nm. The insets illustrates the optical
micrograph and dark field images of the monolayer; scale bar 25 \textgreek{m}m.}
\end{figure}

\clearpage{}

\subsubsection{~~Semiconductor model:}

The O\textquoteright Donnel model gives the following values of {\footnotesize{}$E_{g}(0)=$2.07
eV (2.02 eV), S=1.78 (2.0) and $\left\langle \hbar\omega\right\rangle =$25
meV (33 meV) for }$X^{0}${\footnotesize{} (}$X^{+}${\footnotesize{})}
correspondingly. The different values of $\left\langle \hbar\omega\right\rangle $
for $X^{0}$ and $X^{+}$ might come from the fact that our temperature
range covers only the decaying part of T (for T below 77 K the peak
position reaches a quasi-steady state and both resonances barely move).
The extracted parameters agree with previously reported values \citesupp{hanbicki_measurement_2015_s}.
It is worth noticing that the $\left\langle \hbar\omega\right\rangle $,
although different for exciton and trion, is similar to the energy
of the LA phonon \citesupp{jin_intrinsic_2014_s}. The $S$ value is rather
homogeneous among TMD monolayer and increases as the number of layers
is decreased \citesupp{arora_excitonic_2015_s} this implies a strong electron-phonon
coupling caused by confinement owned by the monolayer.
\begin{figure}[H]
\centering
\includegraphics[width=10cm]{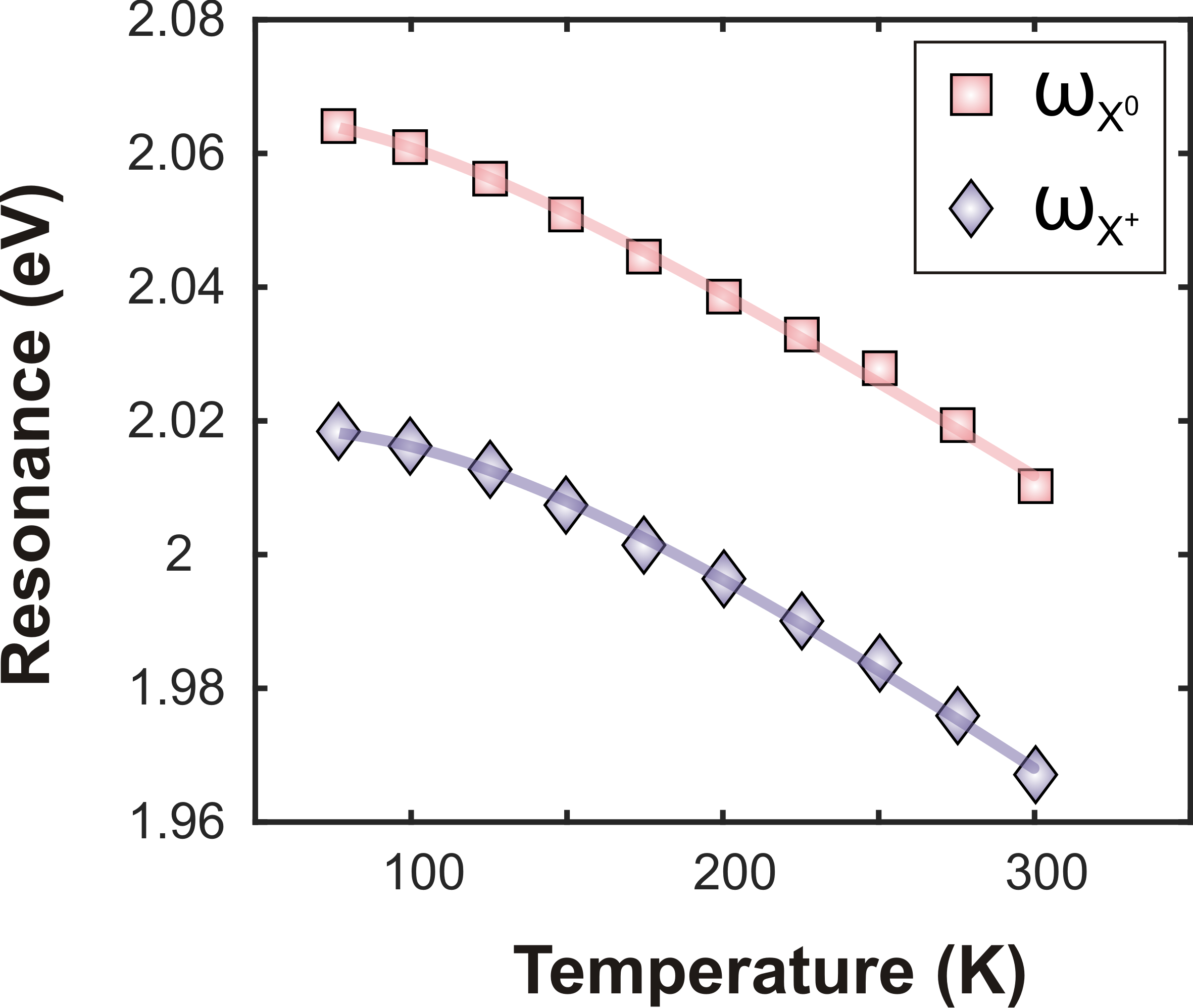}
\caption{Uncoupled exciton and trion vs temperature. Uncoupled exciton and
trion resonances extracted from the PL data in Fig. S3a. O\textquoteright Donnel
fits are shown by solid lines. The O\textquoteright Donnel model gives
the following values of {\footnotesize{}$E_{g}(0)=$2.07 eV (2.02
eV), S=1.78 (2.0) and $\left\langle \hbar\omega\right\rangle =$25
meV (33 meV) for }$X^{0}${\footnotesize{} (}$X^{+}${\footnotesize{})}
correspondingly.}
\end{figure}

\clearpage{}

\subsubsection{~~Rabi Oscillations:}

\begin{figure}[H]
\centering
\includegraphics[width=12.75cm]{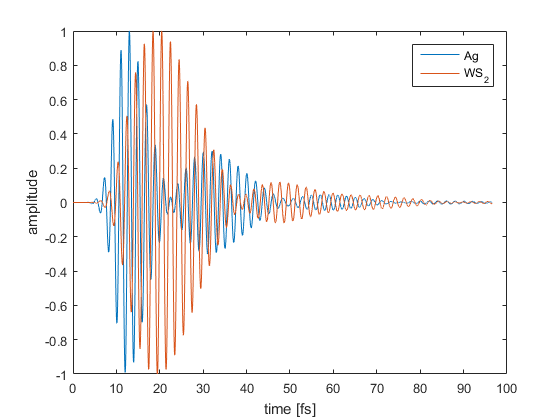}

\caption{Visualisation of Rabi oscillations using FDTD. Field amplitudes as
a function of time. Note ultrafast energy transfer between metal and
WS\protect\textsubscript{2} corresponding to several Rabi cycles.To
monitor the temporal evolution of the excited we track the amplitude
of the electric field of the plasmon of the Ag nanoprism and the auxiliary
electric field in the WS\protect\textsubscript{2}, which is a measure
of the amount of energy stored in the 2D material in the form of excitons.}
\end{figure}

\clearpage{}

\subsubsection{~~Plasmon-Exciton-Trion Hamiltonian:}

To understand the dynamics of the coupled system we analyze the eigen
problem of the Hamiltonian describing the coupled system. We use the
following basis: $\left|1,g,g\right\rangle $, $\left|0,e,g\right\rangle $,
$\left|0,g,e\right\rangle $ , where the corresponding symbols refer
to plasmon, exciton and trion excitations. These notations lead to
the following Hamiltonian:

\[
\frac{\hat{H}}{\hbar}=\left(\begin{array}{ccc}
\omega_{pl}-i\gamma_{pl} & g_{X^{0}} & g_{X^{+}}\\
g_{X^{0}} & \omega_{X^{0}}-i\gamma_{X^{0}} & 0\\
g_{X^{+}} & 0 & \omega_{X^{+}}-i\gamma_{X^{+}}
\end{array}\right)
\]

Here $\omega_{pl}$, $\omega_{X^{0}}$, $\omega_{X^{+}}$ and $\gamma_{pl}$,
$\gamma_{X^{0}}$, $\gamma_{X^{+}}$ are resonance frequencies and
dissipation rates for plasmon, exciton and trion respectively. The
coupling strength between plasmon-exciton and plasmon-trion are denoted
as $g_{X^{0}}$ and $g_{X^{+}}$ correspondingly. Possible exciton-exciton,
trion-trion and exciton-trion interactions are neglected in this model.

There are three hybrid states (lower \textendash{} LP, middle \textendash{}
MP and upper \textendash{} UP polaritons) arising from solving the
eigenvalues of the system. These hybrid states may be expressed as
linear combinations of the original plasmon, exciton and trion excitations
with complex valued coefficients. The absolute square values of the
coefficients give insight on the plasmonic, excitonic and trionic
contributions to each polariton branch.

We solve the corresponding eigen problem numerically. The contributions
from each constituent system to the polariton branches turn out to
be highly dependent on the coupling strength and the detuning. By
studying the results when combined strong coupling $(2\sqrt{g_{X^{0}}^{2}+g_{X^{+}}^{2}}\gtrsim\gamma_{pl})$
is just reached the MP contains a photonic component close to \ensuremath{\sim}20\%
(see Fig. S7a). Though in the very strong coupling case $(2g_{X^{0}}>\gamma_{pl}$,
$\gamma_{X^{0}})$ where the coupling also exceeds the detuning between
exciton and trion states, the plasmonic component nearly disappears
and thus the middle polariton becomes dark (see Fig. S7b). Since this
is not observed in the experiment (see Main text), we conclude that
our system adopts the just reached combined strong coupling regime
in agreement with Fig. S7a.

\begin{figure}[H]
\centering
\includegraphics[width=15cm]{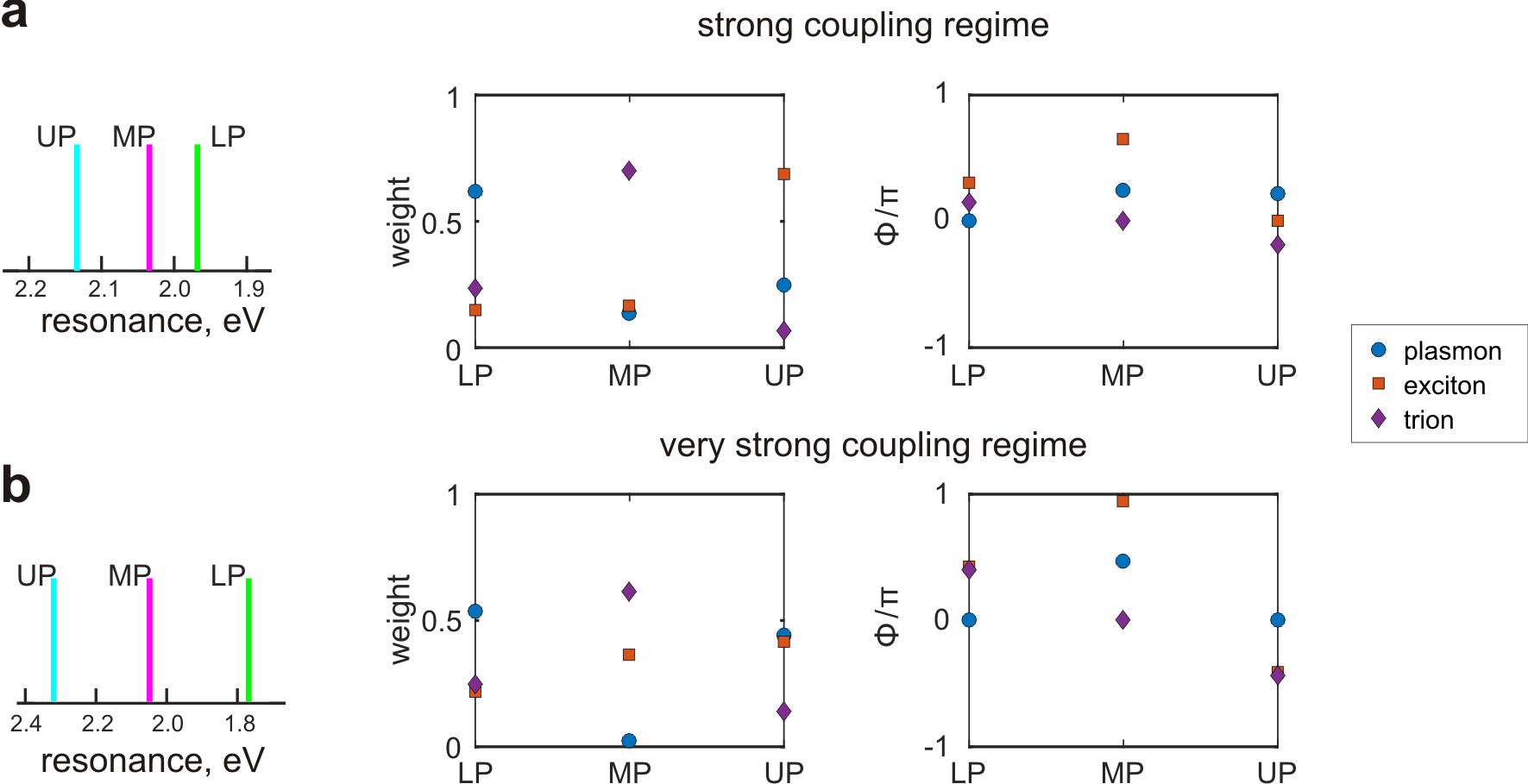}
\caption{Intermediate and strong coupling regimes in 3\texttimes 3 Hamiltonian.
(a) Intermediate coupling regime is modelled by setting the following
parameter values: $\omega_{pl}=$2.02 eV, $\omega_{X^{0}}=$ 2.10
eV, $\omega_{X^{+}}=$ 2.01 eV; $\gamma_{pl}=$ 200 meV, $\gamma_{X^{0}}=$
50 meV, $\gamma_{X^{+}}=$ 60 meV; $g_{X^{0}}=$ 80 meV, $g_{X^{+}}=$
64 meV. These values are extracted from the experimental data using
the coupled oscillator model. Note that MP has approximately 20\%
plasmon contribution meaning that MP is bright. (b) Strong coupling
regime is modelled by setting the following parameter values: $\omega_{pl}=$2.02
eV, $\omega_{X^{0}}=$ 2.10 eV, $\omega_{X^{+}}=$2.01 eV; $\gamma_{pl}=$
200 meV, $\gamma_{X^{0}}=$ 50 meV, $\gamma_{X^{+}}=$ 60 meV; $g_{X^{0}}=$220
meV, $g_{X^{+}}=$176 meV. Note that MP has nearly zero plasmon contribution
meaning that MP is essentially dark.}

\end{figure}
\clearpage{}

\nocitesupp{*}
\bibliographystylesupp{apsrev4-1}
\bibliographysupp{SI}

\end{document}